\hfuzz 9.5pt
\hsize=12.5cm
\vsize=19.2cm
\parindent=0.5cm
\parskip=0pt
\baselineskip=12pt
\topskip=12pt

\magnification=1200     
\baselineskip=16pt

\def\nl{\par\noindent}

\pretolerance=10000

\nopagenumbers
\headline={\ifnum \pageno=1 \hfil \else\hss\hskip44.8pt\tenrm- \folio\
-\hss\fi {\tt
CBPF-NF-063/96}}

\font\ninerm=cmr9

%\font\csc=cmcsc

\def\({\c c}
\def\|{\'\i}

\centerline {{\bf $\theta$-Vacua in the Light-Front Quantized  
Schwinger Model} 
\footnote{$^{\dagger}$}{\ninerm {\it Invited talk given at the 
 {\it First US-Brasilian Workshop on Topology, Geometry, and
Physics}, UNICAMP, Campinas, SP, Brasil, July '96} 
{\sl and at the} {\it Intl. 
Workshop on Light-Front Quantization and Non-Perturbative QCD}, {\sl Iowa
State University, Ames, Iowa, USA, March '96. {\rm To be published in the 
Proceedings of the US-Brasilian Workshop.} } }}

\medskip
\bigskip
\centerline {Prem P. Srivastava\footnote{$^{\star}$}
 {\ninerm {E-Mail:  }  
prem@vax.fis.uerj.br and lafexsu1.lafex.cbpf.br
} }
\medskip
\centerline {\it  Instituto de F\|sica, Universidade do Estado de  
Rio de Janeiro, }
\centerline {\it Rua S\~ao Francisco Xavier 524,  
Rio de Janeiro, RJ, 20550-013, Brasil. }

\centerline {\sl and}

\centerline {\it Centro Brasileiro de Pesquisas F\|siscas-CBPF/LAFEX}
\centerline {\it Rua Xavier Sigaud, 150, Rio de Janeiro, RJ, 22290, Brasil}
\vskip 1.3cm
\centerline {\bf Abstract}
\medskip

{\leftskip 30pt\rightskip 30pt {\tenrm  The  light-front (LF)  
 quantization of the 
bosonized Schwinger model is discussed in the {\it continuum
formulation}. The proposal, successfully used  earlier for describing the 
spontaneous symmetry breaking (SSB) on the LF,  
  of separating first the 
scalar field  into the dynamical 
condensate and the fluctuation fields before employing the standard Dirac 
method works here as well. The condensate variable, however, is now shown
to be a q-number operator in contrast to the case of SSB where it 
was shown to be a c-number or a background field. 
The condensate or $\theta$-vacua emerge straightforwardly 
together with their continuum normalization which avoids the 
violation of the cluster decomposition property in the theory. Some
topics on the {\it front form } theory are summarized in the
Appendices and attention is drawn to the fact that the theory quantized,
say, at equal $x^{+}$  seems already to carry information on equal $x^{-}$ 
commutators as well. }\par}


\nl IF-UERJ 034/96
\nl September 1996

\vfill\eject
\bigskip
\nl {\bf 1. \quad Introduction}
\medskip

Dirac [1] in 1949 pointed out the advantages of studying the relativistic 
quantum dynamics of  physical system on the hyperplanes of the
light-front: $x^{0}+x^{3}=const.$, the {\it front form}. The 
LF or light-cone coordinates, in place of  $\,(x^{0},x^{1},x^{2},x^{3})$,  are 
then convenient to use; they are 
defined by 
$(x^{+},x^{-},x^{\perp})$ where $x^{\pm}=(x^{0}{\pm} x^{3})
/{\sqrt 2}=x_{\mp}$ and   $ x^{\perp}= 
(x^{1}, x^{2})$. The metric tensor for the indices  $\mu= +,-,1,2$ 
may be read from the Lorentz invariant expression 
%is  $g^{++}=g^{--}=g^{12}=g^{21}=0, \;
%g^{+-}=g^{-+}=-g^{11}=-g^{22}=1$ and 
$A^{\mu}B_{\mu}=g_{\mu\nu}A^{\mu}B^{\nu}=g^{\mu\nu}A_{\mu}B_{\nu}= 
A^{+}B^{-}+A^{-}B^{+}-A^{\perp}.B^{\perp}$. 
We make the {\it convention} to regard \footnote{$^{\ddag}$}{\ninerm  
We can of course make the convention with the role of $x^{+}$ and 
$x^{-}$ interchanged. In Appendix D we illustrate by an 
example how 
the  equal-$x^{+}$ quantized theory does seem to contain 
information on the equal-$x^{-}$ commutators as well.}
$x^{+}\equiv \tau$ as the 
LF {\sl time} coordinate while $x^{-}\equiv x$ is the longitudinal 
spatial coordinate and we study the evolution  in $\tau$ 
of the dynamical system. 
The LF components of any four-vector or 
tensor are similarly defined. We note  that the separation 
of two points $x$ and $y$ on equal-$\tau$ plane 
is also spacelike. It becomes lightlike when $x^{\perp}
=y^{\perp}$ but, unlike in the equal-time case, the points 
need not be coincident since ($x^{-}-y^{-}$) may take any value. The  
{\it microcausality principle} leads to locality requirement 
only in $x^{\perp}$ and the appearence 
of any nonlocality in the longitudinal coordinate in the theory would
not be unexpected [2].

The transformation from the conventional to LF 
components is, however, {\it not  a Lorentz transformation} and the 
structure of the {\it LF  phase space} is notably different when 
compared with the conventional phase space. For example,  
the  momentum 
four-vector  is $(k^{-}, k^{+}, k^{\perp})$ where 
$k^{-}$ is the LF energy while $k^{\perp}$ and  $k^{+}$ 
indicates the transverse and the longitudinal components of the  
momentum. A massive particle on the mass shell, 
$\,k^{-}=(m^{2}+{k^{\perp}}^{2})/(2k^{+})$, 
 has  positive definite values for $k^{\pm}$ in contrast to 
$-\infty \leq k^{1,2,3}\leq \infty$ for the usual  components. 
An immediate consequence is that the  {\it vacuum} in the LF 
{\it quantized theory } may become  simpler than the one in the 
conventional (equal-time) theory and in many cases 
the interacting theory vacuum on the LF 
may be the same as the perturbation theory vacuum. 
For example, the conservation of the total  
longitudinal momentum would not permit the excitations of
particle-antiparticle pairs by the LF vacuum (having  $\sum
k^{+}=0$). The SSB on the LF, for example, 
is described [2] 
in a way different  from the conventional one even though the 
physical outcome, as expected, is the same (Apppendix C).

An important advantge pointed out by Dirac is that {\it seven} out of the ten   
Poincar\'e generators are  {\it kinematical } on the LF while in the  
conventional theory  constructed on the hyperplanes $x^{0}=const$., 
the {\it instant form}, only {\it six} have this property. Also the 
notions of spin on the LF for massive and massless particles seem to get
unified (Appendices  A and B). 

Another notable feature of a relativistic theory in the {\it front
form}  is 
that it gives rise to a singular Lagrangian, e.g., 
a constrained dynamical system [3]. It leads in
general to a reduction in the number of independent field operators 
on the corresponding  phase space. The
vacuum structure may then become more tractable and the 
computation of  the physical observables  may become simpler.  It is 
illustrated below by the detailed study of the Schwinger model [4]. 
In the conventional framework, 
for example, the QCD vacuum is quite complex 
due to the {\sl infrared slavery} and  it contains  
also the gluonic and fermionic condensates. 
There seems also to exist  a contradiction between the 
Standard Quark Model and the QCD containing a sea of partons (quarks,
anti-quarks and gluons); the {\it front form} 
theory may throw  light on such problems. 

The LF field theory was rediscovered in 1966 by Weinberg [5] in his 
Feynman rules adapted for infinite momentum frame. Latter  it was 
demonstrated [6] that these  rules correspond to 
the quantization of the theory on the LF. 
The recent revival [7-9] of the interest in LF quantization 
owes, say, to the difficulties encountered in the computation of 
nonperturbative effects in the {\it instant form} QCD 
or in the study of the relativistic bound states of light fermions
[8,7] which can not be handled, say, by the Lattice gauge theory. 
The LF coordinates have proved also very useful in the study of the theories 
of (super-) strings and membranes as well. 

The  purpose of the present work is to show 
how the  quantization of the massless 
Schwinger model on the LF leads in a straightforward way  to the 
$\theta$-vacua  known to emerge [10,11] in the  {\it instant form} 
theory.  
The important feature   of the 
continuum normalization of these states 
arises  naturally on the LF. This is in contrast to the 
discussions in the conventional 
Lorentz  or the Coulomb gauge framework  where we  
have to invoke arguments to  impose it, so as to avoid [10-12] 
the violation of the the cluster decomposition property.  
The discussion on the LF is more transparent due to a 
reduced number of  independent fields.  It is obtained 
here by making  use of the method proposed  
earlier in connection with the  {\it front form} description 
 of the SSB (and the tree level 
Higgs mechanism) [13,2]. 
The  scalar field (of the equivalent bosonized 
Schwinger model) is 
separated, based on physical considerations, into the dynamical 
bosonic condensate and the quantum fluctuation fields. The 
Dirac procedure [3] is then followed in order to 
construct the Hamiltonian formulation and the quantized theory.

After a brief discussion in Sec. 2 of the Schwinger model its 
bosonized version  is quantized  on the LF. The 
{\it condensate} or $\theta$-vacua are described in Sec. 3 while the
conclusions are summarized in Sec. 4. Appendices contain some 
material related to the {\it front form} field theory.

\bigskip
\nl {\bf 2.\quad LF Quantization of the  Schwinger Model}
\medskip

The  field theory model is simply  the two dimensional massless quantum
electrodynamics  with the Lagrangian 

$${\cal L}=\bar\psi\,i\gamma^{\mu}\partial_{\mu}\psi-{1\over 4}F^{\mu\nu}
F_{\mu\nu}-e\bar\psi\,\gamma^{\mu}\psi A_{\mu},\eqno(1)$$

\nl where $\psi$ is a two-component spinor field\footnote{$^\dagger$}
{\ninerm  $\gamma^{0}=\sigma_{1}, \gamma^{1}=i\sigma_{2}, 
\gamma_{5}=-\sigma_{3}, \eta^{00}=-\eta^{11}=1,
\epsilon^{-+}=\epsilon^{01}=+1$, and $\gamma_{5}\gamma^{\mu}=\epsilon^{\mu\nu}
\gamma_{\nu}$ etc.}  and $ A_{\mu}$ is the $U(1)$ gauge field.  
The Lagrangian  is  
 invariant under the global 
$U(1)_{5}$ chiral transformations $\psi\to exp(i\gamma_{5}
\alpha)\,\psi $ apart from under the usual $U(1)$ gauge
transformations. The model is exactly solvable and its 
physical spectrum consists solely of  massive vector field [4]. 
This was made explicit by Lowenstein and Swieca [10,11] through their 
operator solution of the model in the {\it instant form} framework 
in the Lorentz gauge and where  the complex structure of the 
ground state was also studied. 
It has also been studied in the Coulomb gauge [12] and on the
light-cone in the discretized formulation [14] 
where there are many subtleties in the fermionic
version (1). 
 We will work here directly 
in {\it the continuum formulation} (Appendix C) on
the LF so that the spurious finite volume effects get  automatically 
suppressed. For 
the purposes of studying the vacuum structure it is also convenient
to study the {\it equivalent} bosnized version. 

The LF coordinates are now $(x^{+},x^{-})$ where $x^{\pm}=(x^{0}{\pm} x^{1})
/{\sqrt 2}=x_{\mp}$.  
The conserved Noether currents defined by 
$j_{\mu}=
\bar\psi\gamma_{\mu}\psi$ and $j_{{5}{\mu}}=
\bar\psi\gamma_{5}\gamma_{\mu}\psi$ satisfy the 
relation $j_{5}^{\mu}=\epsilon^{\mu\nu} j_{\nu}$ which is 
particular to the two dimensional theory. Written explicitly, 
$\;j_{5}^{+}=-j^{+}
 =-{\sqrt 2}{\psi_{2}}^{\star}\psi_{2},\,\; j_{5}^{-}=+j^{-} 
={\sqrt2}{\psi_{1}}^{\star}\psi_{1}$. The classical conservation 
of both the current  leads to 
$j^{+}=j^{+}(x^{-})$ and $j^{-}=j^{-}(x^{+})$  like  
in the case of the free Dirac field. 
In the quantized theory, however, 
 the currents must be defined, say, by the introduction of the 
of the point-splitting of 
operators. When the  gauge coupling is also 
present we are required to  construct   gauge
invariant currents together with the point-splitting.  
The divergence of the axial current is then found  anomalous: 
\quad $\partial_{\mu}j_{5}^{\mu}
=({e/({2\pi})})\epsilon_{\mu\nu}F^{\mu\nu}$ and the eqs. of motion
lead to $(\partial_{\mu}\partial^{\mu}+{e^{2}/{\pi}})\epsilon^{\mu
\nu}F_{\mu\nu}=0$ implying 
dynamical mass generation [4] for the gauge field in the quantized
theory. The same results may also be obtained by employing functional 
integal methods [11].

The abelian gauge theory under consideration has been extensively studied 
and its  exact solvability  [4,10] 
derives from the remarkable 
property of one-dimensional fermion systems, viz, that they can 
equivalently be described in terms of canonical one-dimensional boson
fields [15].  The operator 
solution [10] mentioned above in a sense amounts to bosonization. 
 Some of the 
relevant correspondences in abelian bosonization  
are \quad  $\bar\psi\psi=K :\cos 2\sqrt{\pi}\,
\phi:,\,\bar\psi\gamma_{5}\psi=K :\sin 2\sqrt{\pi}\,\phi:,\,
\bar\psi\gamma_{5}\gamma_{\mu}\psi=\partial_{\mu}\phi/\sqrt{\pi},\,
\bar\psi\gamma_{\mu}\psi=\epsilon_{\mu\nu}\partial^{\nu}\phi/\sqrt{\pi},\,
\bar\psi\,i\gamma.\partial\psi=
{1\over2}\partial_{\mu}\phi\partial^{\mu}\phi\,$ where  $\phi$ is a
bosonic scalar field and   $K$ is a constant. The fermionic
condensate $<\bar\psi\psi>_{0}$, for example,  may then be expressed 
in terms of the value of the bosonic condensate. The bosonized 
theory can also be constructed 
with the use of the functional integral method [11]. 
The original fermionic and the bosonized theories 
are {\it equivalent} in the sense 
that they have the same current 
commutation relations and the energy-momentum tensor is the same when
expressed in terms of the currents. 

% based on the [Cabra, Rothe and Schaposnik]
%decoupling of the interacting fermions which become free 
% and which in turn be bosonized in terms of the Wess-Zumino fields. 

The Lagrangian density 
of the {\it bosonized} massless Schwinger Model takes the following form 

$${\cal L}= {1\over2}\partial_{\mu}\phi\partial^{\mu}\phi-g
A_{\mu}\epsilon^{\mu\nu}\partial_{\nu}\phi-
{1\over4}F^{\mu\nu}F_{\mu\nu},\eqno(2)$$

\nl where $g={e/\sqrt{\pi}}$. It carries in it
all the symmetries of the original fermionic model including the 
information on the mass generation.  
Under  the $U(1)$ gauge field transforamtion the scalar field 
is left invariant 
while under the 
$U_{5}(1)$ chiral transformations, in view of the 
correspondences above, it suffers a translation  by a 
constant.  This is 
crucial in obtaining the so called $\theta$- or {\it
condensate} vacua. 
The quantization of the bosonized theory 
would allow us to describe the vacuum structure of the original theory and
to compute, say, the fermionic condensate.

We first make  the {\it separation}, proposed in Ref. [13], 
in  the scalar field (a generalized function) 
:\quad  $\phi(\tau,x^{-})= \omega (\tau)+\varphi(
\tau,x^{-})$,  where  $\omega(\tau) $  is the (
dynamical) bosonic {\it condensate} variable   and the field $\varphi$   
represents the  quantum  
fluctuations. This 
%above the condensate. 
%Such a separation 
enabled us to give [13] a 
description on the LF of the SSB and (tree
level) Higgs mechanism and where the variable  $\omega$ was shown 
to be a c-number, e.g., a background field. 
%It was also shown in that discussion 
%that the LF Hamiltonian is non-local and that in the LF quantized scalar
%theory  $\omega$ is a c-number (e.g., a background field). 
In the Schwinger model, on the contrary, it will be shown
below to turn  out as a  q-number operator and its eigenvalues would 
label the vacuum states. 
We will set $\int dx^{-} 
\varphi(x^{+},x^{-})=0$ so that the entire zero-momentum mode of
$\phi$ is 
represented by the condensate variable. 
The {\it chiral transformations} would be defined [12] as:\quad
$\omega\to \omega+const., \; \varphi\to \varphi$,  and $A_{\mu}\to A_{\mu}$ 
so that  the boundary conditions at infinity on $\varphi$ are left 
unaltered (see Sec. 3). The bosonized Lagrangian   then becomes 

$$L= \int_{-R/2}^{R/2} dx^{-}\Bigl[{\dot\varphi}\varphi' +g(A_{+}\varphi'
-A_{-}{\dot\varphi})+{1\over 2}({\dot A}_{-}-{A_{+}}')^{2}\Bigr]-
g{\dot\omega}\int_{-R/2}^{R/2} dx^{-} A_{-}(\tau,x^{-})\eqno (3) $$

\nl Here $R\to\infty$, an overdot (a prime) indicates the partial derivative 
with respect to $x^{+}\equiv \tau \; (x^{-}\equiv x)$, 
and $\varphi$ is assumed to satisfy the conditions required for 
the existence of the Fourier transform in the spatial variable $x^{-}$. 

The last term of (3) shows that the 
light-cone gauge $A_{-}=0$ is not convenient to adopt in the present
case. It is suggested and in fact will be shown below 
that the zero-momentum mode of $A_{-}$, viz,  $h(\tau)\equiv 
\int dx^{-} A_{-}$ and $\omega$ 
form a canonically conjugate pair. We would instead adopt the
gauge [16]  $\partial_{-} A_{-}=0$ which is 
shown to be  accessible on the phase space. 
The  canonical momenta defined  from (3) are 

$$\eqalign {\pi={{{\delta {\cal L}}\over {\delta {\dot\varphi}}}} &=
\varphi'-g A_{-}\cr
E^{+}&={{{\delta {\cal L}}\over {\delta {\dot A}_{+} }}}=0\cr 
E^{-}&={{{\delta {\cal L}}\over {\delta {\dot A}_{-} }}}=({\dot A}_{-}
-{A_{+}}')\cr
p&={{\partial L}\over {\partial \dot\omega}}=-g h(\tau)\equiv -\pi_{\omega}\cr 
}.\eqno(4)$$

\nl The primary constraints [3] are thus $\chi\equiv 
\pi- \varphi'+g A_{-}\approx 0$, $E^{+}\approx 0$, and $T(\tau)\equiv
(-\pi_{\omega} + g h)\approx 0$, 
where $\approx$ indicates the weak equality [3]. 
The canonical Hamiltonian is found to be 

$$H_{c}=\int dx^{-} \Bigl[{1\over 2} {E^{-}}^{2}+E^{-} {A_{+}}'
-g A_{+} \varphi'\Bigr ].\eqno(5) $$

\nl and we take  the preliminary [3] Hamiltonian  as 

$$H'=H_{c}+\int dx^{-} \Bigl[u_{+}E^{+}+ u \chi \Bigr ] + \lambda(\tau)
T(\tau), \eqno (6) $$

\nl where $u_{+},u,$ and $\lambda$ are the Lagrange multiplier
fields. We assume initially the standard equal-$\tau$ Poisson brackets: 
$\{\pi_{\omega}(\tau),\omega(\tau)\}=+1$, $\{E^{\mu}(\tau,x^{-}),A_{\nu}(\tau,
y^{-}) \}=-\delta_{\nu}^{\mu}\delta (x^{-}-y^{-})$,  
$\{\pi(\tau,x^{-}),\varphi(\tau,y^{-}) \}=-\delta(x^{-}-y^{-})$ etc. from
which it follows, for example, that  $\{E^{\mu}(\tau,x^{-}),h(\tau)
 \}=-\delta_{-}^{\mu}$. 
On requiring the persistency of the constraints in $\tau$ employing 
$df/d{\tau}=\{f,H'\}+\partial f/\partial {\tau}\,$ we derive the
secondary constraints $\;K(\tau)\equiv \int dx^{-} [ E^{-}+A_{+}'
 ]\approx 0$ and $\Omega\equiv
\partial_{-}(E^{-}+g\varphi)\approx 0$. We go over now to an extended
Hamiltonian including these constraints as well and repeat the
procedure. No new  constraints are shown to be generated since  only
the equations which would 
determine the Lagrange multiplier fields are left. 
The constraints $\Omega$ and $E^{+}$ are easily shown 
to be first class [3] while 
$T$, $K$, and $\chi$ are second class [3]. From the eqs. of motion we 
show  that we may determine the multiplier fields such that 
 $\,\partial_{-}A_{-}\approx0\,$ and $\,(E^{-}+A_{+}')\approx0\,$
along with their persistency conditions are satisfied. 
We may  thus add to the theory these ones as the ( external) gauge-fixing 
constraints,  corresponding 
to the two first class constraints found above. 
% (which generate gauge transformations on the phase space). 
The whole set of constraints then becomes second 
class and we may proceed to construct Dirac brackets [3], which 
would replace the Poisson ones, such that the weak equalities 
may be replaced by the strong equalities (even) inside them. 
%$\,df/d{\tau}=\{f,H'\}_{D}+\partial f/\partial {\tau}\,$ determine
%the Hamilton's equations. 

It is straightforward to construct the Dirac bracket iteratively.  
We first handle the pair $T\approx0$, $K\approx0$ with 
$\{T,K\}\approx gR$, $\{T,T\}\approx 0$, and $\{K,K\}\approx 0$.  
The corresponding 
modified equal-$\tau$  bracket is easily constructed\footnote{$^{\ddag}$}{
We make the convention that the 
first variable in an equal-$\tau$ bracket $\{f,k\}$ refers to
longitudinal coordinate $x^{-}$ while the second  one  to 
$ y^{-}$. We remind that we are working 
{\it  in the continuum formulation} and 
that $R\equiv\int _{-R/2}^{R/2}dx\to \infty $ at the end of the
computation.}

$$\{f,k\}_{1}=\{f,k\}+{1\over {gR}}\Bigl[\{f,T\}\{K,k\}-
\{f,K\}\{T,k\}\Bigr],\eqno(7)$$

\nl We verify, for example, $\{f,T\}_{1}\equiv 0$, $\{K,f\}_{1}\equiv
0$ etc.. The modified bracket differs from the Poisson one only when the 
variables $\omega, E^{-},
E^{+}$, or $A_{-}$ are to be found in the functionals 
$f$ and $k$. For example, 
$\{\omega, h\}_{1}=-(1/g)$, $\{\omega, E^{+}\}_{1}=0$, $\{E^{-}, A_{-}\}_{1}
=\{E^{-},A_{-}\}+ (1/R)$, $\{f, \chi\}_{1}=\{f,\chi\}-(1/R)\{f,T\}$, 
$\{\chi,\chi\}_{1}=\{\chi,\chi\}=
-2\partial_{-}^{x}\delta(x^{-}-y^{-})$.

A further modification 

$$\{f,k\}_{2}= \{f,k\}_{1}+{1\over 4} \int\int du dv \; \{f,\chi\}_{1}
\epsilon(u-v) \{\chi,k\}_{1},\eqno(8)$$

\nl allows us to set $\chi=0$ also as strong equality. We find, for 
example,  $\{\omega,\omega\}_{2}=0$, $\{E^{-},E^{-}\}_{2}= 
(-g^{2}/4) [ \epsilon(x-y)+(1/R)\int du \;
\{\epsilon(u-x)-\epsilon(u-y)\} ]$, $\{\omega,E^{-}\}_{2}= 
-(g/(4R))\int du\, \epsilon (u-y) $, $\{\varphi,\pi\}_{2}=(1/2)
\delta(x-y)$, $\{\varphi,\varphi\}_{2}=(-1/4)\epsilon(x-y)$, 
$\{\varphi,E^{-}\}_{2}=(g/4)\epsilon(x-y)$,   $\{E^{+},\Omega\}_{2}
=0$, $\{\Omega,\Omega\}_{2}=0$ etc., and  that $E^{+}$ and $\Omega$
continue to remain first class even with respect to 
$\{,\}_{2}$. 

Next the   {\it gauge-fixing} constraint $\chi_{3}\equiv
\partial_{-}A_{-}\approx 0$ along with $\Omega\approx 0$  are 
implemented. 
%which is shown to be accessilble as seen
%form the eqs. of motion derived using now the brackets $\{,\}_{2}$.
The constraint matrix has elements $\{\chi_{3},\chi_{3}\}_{2}=0$, 
$\{\chi_{3},\Omega\}_{2}=-\partial_{x}^{2}\delta(x-y)$,  
 and  $\{\Omega,\Omega\}_{2}=0$ and its inverse is easily 
found to be $i\sigma_{2}\,\partial_{x}^{-2}
\delta(x-y)$. Hence we define

$$\{f,k\}_{3}=\{f,k\}_{2}+\int du\; \Bigl 
[\{f,A_{-}\}_{2}\{E^{-}+g\varphi,k\}_{2}
-\{f,E^{-}+g\varphi\}_{2}\{A_{-},k\}_{2}\Bigr]\eqno(9).$$

\nl We find  $\{A_{-},\omega\}_{3}=1/(gR)$ and 
%$\{A_{-},f\}_{3}=(1/R) \{h(\tau),f\}_{1}$ 
$\{A_{-},f\}_{3}\to 0$ when $R\to\infty$ for $f\neq \omega$. 
Some of the others are $\{\omega,E^{-}\}_{3}= 
(-g/(4R))\int du \,\epsilon(u-y)$, $\{h,E^{-}\}_{3}=0$, 
$\{\omega,h\}_{3}=-1/g$, 
$\{E^{-},E^{-}\}_{3}=(-g^{2}/4)\epsilon(x-y)$. 

The remaining first class constraint $E^{+}\approx0$  may be
taken care of by adding still another {\it gauge-fixing} constraint  
$\Phi\equiv E^{-}+A_{+}'\approx0$. The two dimensional constraint
matrix  
$C(x,y)$ here has the elements: $\;C_{11}=\{E^{+},E^{+}\}_{3}=0$, 
$C_{12}=C_{21}= \{E^{+},\Phi\}=\{E^{+},A_{+}'\}=\partial_{x}\delta(x-y)$,
 $C_{22}=\{\Phi,\Phi\}_{3}=-g^{2}\epsilon(x-y)/4$.  An inverse
matrix $C^{-1}$ is shown to have the elements $C_{12}^{-1}=C_{21}^{-1}=
(1/2)\epsilon(x-y)$ and  $C_{22}^{-1}=0$ while $C_{11}^{-1}$
satisfies $\partial_{x}^{2}
C_{11}^{-1}(x,y)=(g^{2}/4)\epsilon(x-y)$. The final Dirac bracket
which implements all the constraints is thus constructed to be  

 $$\eqalign {\{f,k\}_{D}&=\{f,k\}_{3}\cr
& \qquad -{1\over 2}\int \int du dv\; \Bigl 
[\{f,\Phi\}_{3}\epsilon(u-v)\{E^{+},k\}_{3}
-\{f,E^{+}\}_{3}\epsilon(u-v)\{\Phi,k\}_{3}\cr
& \qquad +2 \{f,E^{+}\}_{3}\, 
C_{11}^{-1}(u,v)\,\{E^{+},k\}_{3}\Bigr]\cr }\eqno(10)$$

\nl We find $\{E^{-},E^{-}\}_{D}=-g^2\epsilon(x-y)/4$, 
$\{\varphi,\varphi\}_{D}=-\epsilon(x-y)/4$, 
$\{A_{-},k\}_{D}=0$ for $k\not=\omega$, $\{A_{-},\omega\}_{D}=1/(gR)$, 
$\{\pi_{\omega},\omega\}_{D}=1$, 
%where $\pi_{\omega}(\tau)\equiv g
%h(\tau)$,  
$\{A_{+},A_{+}\}_{D}=C_{11}^{-1}(x,y) $, and $\{\omega,E^{-}\}_{D}=
-g\int du\,\epsilon(u-y)/(4R)$ among the others.  All the constraints
can now be written as strong equalities and 
we are left behind with the independant variables 
$\varphi $, $\omega$, $\pi_{\omega}$ and 
$E^{-}=-\partial_{-} A_{+}=-g\varphi$. The 
Hamiltonian density is effectively given 
by ${\cal H}_{D}={E^{-}}^{2}/2 +\partial_{-}
(E^{-}A_{+})$ and  $H_{D}= \int dx \,{E^{-}}^{2}/2= g^2\int dx\,
 \varphi^{2}/2$ corresponding to a free scalar field $\varphi$ of mass
$e^{2}/\pi$. It may be checked that for the self-consistency with the
Lagrangian eqs. 
%(in the gauge $\partial_{\pm}A_{-}=0$, e.g., $A_{-} is a space-time 
%$independent constant), and  which is also symmetrical in the 
%coordinates $\pm$)
we require that $A_{+}$ should satisfy the periodic 
boundary conditions  at infinity in $x^{-}$.

\bigskip

\nl {\bf 3. \quad  $\theta$-Vacua }
\medskip

We discuss now the vacuum state in the LF quantized
theory. The {\it ultimate physical conclusions} would of course coincide
with those following from the {\it instant form} theory.  
The {\it quantized theory} is obtained through 
the correspondence [3] $i\{f,k\}_{D}\to [f,k]$ with  the 
commutators of the field operators. 
We find  $[\omega,\omega]=0$,  $[\pi_{\omega},\omega]=i$, 
$[\omega,E^{-}]=[\omega,\varphi]=[\pi_{\omega},\varphi]=
[\omega,H_{D}]=[\pi_{\omega},H_{D}]=0$, and 
the well known LF commutator $[\varphi,\varphi]=
-i\epsilon(x-y)/4$ which leads to $\,2\pi[j^{+}(\tau,x),j^{+}(\tau,y)]= 
i\partial_{-}\delta(x-y)$, where the right hand side is the 
Schwinger term.  The condensate field $\omega$ 
turns out to be a q-number (operator) in the present model 
in contrast to the case of the LF quantized 
scalar theory where it is shown (Appendix C) to be a c-number. 
This is expected since 
the chiral symmetry in the model 
is realized as translation of the value of the
condensate. The latter must then be represnted by an operator 
$\omega$ in order that its eigenvalues may be shifted by 
another operator. The  $\omega(\tau)$ in our 
discussion has been treated as a dynamical variable 
and we let the Dirac procedure to determine if it 
is  a  c- or q-number in the quantized theory.

The original fermionic Schwinger model is invariant under {\it
global} $U(1)_{5}$ chiral transformations $\psi\to exp(i\alpha \gamma_{5})\,
\psi$, $\bar\psi\psi\to\bar\psi\, exp(i2\alpha \gamma_{5})\,\psi$. 
From $exp(i\alpha\gamma_{5})=(\cos \alpha+i\gamma_{5}\sin
\alpha)$ it is clear that the $U(1)_{5}$ like $U(1)$ is  a compact 
group with $\alpha=\alpha_{0}+2\pi n, \, n=0,\pm 1,\pm 2,..,$ and, 
for example, $n=0$ with 
$0 \leq \alpha_{0} \leq 2\pi$ would enumerate all the distinct 
elements of the group. For $\alpha=n\pi,\;\, n=0,\pm 1,\pm 2,..,$ 
$\bar\psi\psi$ and $\bar\psi\gamma_{5}\psi$ are 
clearly left invariant under chiral transformations. From the correspondence 
$\bar\psi\,\psi\leftrightarrow K :\cos(2\sqrt{\pi}\phi):$ implied 
in the construction of the  bosonized theory it follows that 
the chiral transformation on the bosonic field $\phi$ is realized 
by $\omega \to \omega+\beta
                     /\sqrt{\pi},\;\varphi\to\varphi$
with $\beta=\beta_{0}+n\pi,\,n=0,\pm 1,\pm 2,..,\;0\leq\beta_{0}\leq
\pi $. 
The $\omega$ is therefore an angular (coordinate or) variable  with the 
period $\sqrt\pi$.
%understanding that $\omega$ and $\omega\pm\sqrt\pi$ 
%represent the same number.
 In the quantized field theory the chiral transformations are 
generated by the  unitary operator $Q_{5}(\beta)$
  such that\footnote{$^{\ddag}$} {\ninerm Here $\omega, \varphi$, and 
$\pi_{\omega}$ are field operators while $\beta$, $a$, and $n$ 
are c-no. constants.} $\,Q_{5}^{\dag}(\beta)\,\omega\, Q_{5}(\beta)= 
\omega+\beta/\sqrt{\pi},\,\; Q_{5}^{\dag}(\beta)\,\varphi\, 
Q_{5}(\beta)=\varphi $ and  $Q_{5}(\pi)=Q_{5}(0)$. The commutation 
relations  given above result in  $exp(-ia\pi_{\omega})\,
\varphi\, exp(ia\pi_{\omega})=\varphi,\;exp(-ia\pi_{\omega})\,
\omega\, exp(ia\pi_{\omega})=\omega+a $. The unitary field theory 
operator which generates 
the chirality transformation may then be defined as 
$Q_{5}(\beta)\equiv exp(i\pi_{\omega}\,\beta/\sqrt{\pi})$. 

 In view of the commutation relations above 
the space of states may be built as 
a tensor product, written as $|\varphi)\otimes |n\}$, of a 
conventional Fock space for the massive field $\varphi$ and a space
spanned by the eigenvalues of $\pi_{\omega}$.  From 
$Q_{5}(\pi)=exp(i
\pi_{\omega}\,\sqrt{\pi})=Q_{5}(0)=I$, where $I$ is identity operator, 
 it follows that  $\pi_{\omega}$  has 
discrete eigenvalues :\quad $\;\pi_{\omega}|n\}= 2n\sqrt{\pi}|n\}\;$
where $n=0,\pm 1,\pm 2,..$ and  $\{m|n\}=\delta_{mn}$. 
With the help of  $\,exp(i a\omega)\,\pi_{\omega}\,exp(-ia
\omega)=\pi_{\omega}+a$ we may construct the ladder operators 
$\,d^{\pm}=exp(\mp 2i\sqrt{\pi}\omega)\,$ such that
$d^{\pm}|n\}=\,|(n\pm 1)\}$.  The Hamiltonian  
contains only the field $\varphi$ and it 
commutes with the  chirality operator $Q_{5}(\beta)$ and $d^{\pm}$. 
We have infinite  
degeneracy corresponding to the  {\sl chiral} vacuum states $|0)\otimes
|n\}$ with $n=0,\pm1,\pm2,..$ since $\omega$ and $\pi_{\omega}$ are
absent from the LF Hamiltonian. There are also no 
transitions between these vacua characterized by 
the differnt values of  $n$. In the original Schwinger solution 
the vacuum state was chosen to be the chirally symmetric state 
$|0)\otimes |n=0\}$ which leads to the violation of the cluster
decomposition property. This may be avoided by choosing 
instead the (alternative) vacuum state for  the model 
to be  an eigenstate of the condensate 
operator. The state vectors which are degenerate 
with respect to the eigenvalue $\omega'$ of the condensate operator
are easily constructed\footnote{$^{\S}$}{\ninerm In the coordinate
representation  $\omega|\omega'\}=\omega'|\omega'\}$ and 
$\pi_{\omega}$ is represented by $i\partial/{\partial \omega'}$. 
We recall also the {\it Poisson summation formula} of the distribution 
theory:\quad $\delta(x)={\sum}^{\infty}_{-\infty}e^{i2{\pi}nx}\,$ for
$-1\leq x\leq 1$ and note that $\{\omega'|\pi_{\omega}|\omega''\}=i\partial
\delta(\omega'-\omega'')/\partial \omega' $ etc.}

$$|\omega'\}\otimes|\varphi)={1\over{\pi^{1/4}}}\;
\sum_{n=-\infty}^{\infty}\; e^{2i{\sqrt{\pi}}\,n\,\omega'}\;
|n\}\otimes|\varphi)=e^{i\pi_{\omega}\omega'}|\omega'=0\}
\otimes|\varphi). \eqno(11)$$

\nl where  $\,0\leq \omega'\leq \sqrt\pi\;$ or 
$\;0\leq\theta'\equiv 2\sqrt{\pi}\omega'\leq 2\pi\;$ 
specify the physical values of the condensate. 
These  states  have   
 the continuum normalization 
$\{\omega''|\omega'\}=\delta(\omega''-\omega')$ and it comes out 
naturally in our discussion of the LF quantized theory. 
This is in contrast to the 
arguments  required in the equal-time formulation 
to impose it 
so as to avoid the violation of the cluster decomposition 
property  and other inconsistencies in the theory [10-12].   
The {\it condensate}- or $\theta$-{\it vacuum}  in the theory under 
consideration is the state 
$|\omega'\}\otimes|0) $ with a fixed given value for $\theta'$ 
and we note that $Q_{5}(\beta)|\omega'\}=|\omega'+\beta/\sqrt{\pi}\}$. 
The  vacuum state is 
infinitely degenerate and  there are 
no transitions among the states with different values of $\theta'$.
The chiral symmetry is thus  spontaneously broken. The
corresponding generator $\pi_{\omega}$ also does not annihilate the 
vacuum  state.  
%The vacuum 
%structure of the Schwinger model describes the broken symmetry. 
This feature here is different from the one found in connection with the 
description on the LF of the SSB (and Higgs
mechanism). There the LF 
vacuum is annihilated by all of the symmetry generators (in contrast
to the case of the equal-time formulation)   
and the broken symmetry is manifested in the quantized theory 
Hamiltonian (Appendix C). The fermionic condensate in the Schwinger 
model, for example, may also  be  computed  

$$ \eqalign{  (0|\otimes \{\omega''|: \bar\psi \psi :|\omega'\}\otimes |0)
&= K (0|\otimes \{\omega''| :\cos 2\sqrt{\pi}(\omega+\varphi): 
|\omega'\}\otimes |0)\cr 
&= K (0|\otimes \{\omega''| :\cos (2\sqrt{ \pi} \varphi):
 |\omega'\}\otimes |0)\cos(2\sqrt{ \pi}\omega')\cr
& = K \cos \theta'\;\delta(\omega''-\omega') \cr }\eqno(12)$$ 
\bigskip
\nl {\bf 5. \quad Conclusions}
\medskip

On the LF the $\theta$-vacua in the massless Schwinger model are 
obtained in straightforward fashion on  quantizing   
the equivalent bosonized theory. Self-consistent 
Hamiltonian formulation on the LF may be built by first 
seperating  the scalar field 
into the dynamical condensate and quantum fluctuation fields and then
following the Dirac procedure. The same procedure allowed earlier to 
describe also the SSB and (tree level) Higgs mechanism.  
 The physical results  may get a different description on
the LF compared to the conventional one [13,19]. 
The integrability of $QCD_{2}$  has been 
conjectured [17] recently from the studies in the 
conventional framework employing the equivalent bosonic description.  
It would be interesting to study the vacuua here in the 
{\it front form} theory and also to find an alternative 
(simpler) proof [18] of the conjecture. 
\bigskip
\medskip
\nl {\bf Acknowledgements}
\medskip
Acknowledgements with thanks are due to  James Vary, Stan Brodsky, 
Heinrich Leutwyler, Maria Cristina Abdalla, Elcio Abdalla, and Luiz 
Alvarez-Gaum\'e for 
constructive suggestions. Thanks are due also to Waldyr Rodrigues 
Jr.  
and the  Organizers of the Workshop at the UNICAMP for 
the invitation at the very stimulating interdisciplinary meeting. 
The kind  invitation of James Vary and  the hospitality at the 
Workshop at the Iowa state University,  the 
hospitality at the Theory Division at CERN, where the present 
work started, and the facilities offered at LAFEX 
are gratefully acknowledged. 

\bigskip
\medskip
\nl {\bf Appendix A:  \quad Poincare Generators on the LF} 
\medskip 

The Poincar\'e generators in coordinate system 
$\,(x^{0},x^{1},x^{2},x^{3})$,  satisfy 
$[M_{\mu\nu},P_{\sigma}]=-i(P_{\mu}g_{\nu\sigma}-P_{\nu}g_{\mu\sigma})$ and 
$[M_{\mu\nu},M_{\rho\sigma}]=i(M_{\mu\rho}g_{\nu\sigma}+
M_{\nu\sigma}g_{\mu\rho}-M_{\nu\rho}g_{\mu\sigma}-M_{\mu\sigma}g_{\nu\rho})
\,$  where the metric is 
 $g_{\mu\nu}=diag\,(1,-1,-1,-1)$, $\mu=(0,1,2,3)$ and we take 
 $\epsilon_{0123}=\epsilon_{-+12}=1$. If we define 
 $J_{i}=-(1/2)\epsilon_{ikl}M^{kl}$
and  $K_{i}=M_{0i}$, where   $i,j,k,l=1,2,3$, we find 
$[J_{i},F_{j}]=i\epsilon_{ijk}F_{k}\,$ 
for $\,F_{l}=J_{l},P_{l}$ or  $ K_{l}$ while 
$[K_{i},K_{j}]=-i \epsilon_{ijk}J_{k}, \, [K_{i},P_{l}]=-iP_{0}g_{il},
\, [K_{i},P_{0}]=iP_{i},$ and  $[J_{i},P_{0}]=0$.

 The LF  generators are 
$\,P_{+}, P_{-},P_{1},P_{2}\,$, $M_{12}=-J_{3},\,M_{+-}=-K_{3},\,M_{1-}=
-(K_{1}+J_{2})/{\sqrt 2}\,
\equiv {-B_{1}}, \; M_{2-}=-(K_{2}-J_{1})/{\sqrt 2}
\equiv {-B_{2}}, \;
M_{1+}=-(K_{1}-J_{2})/{\sqrt 2}\equiv -S_{1},\,$ and 
$M_{2+}=-(K_{2}+J_{1})/{\sqrt 2}\equiv -S_{2}\,$. We find 
$[B_{1},B_{2}]=0,  [B_{a},J_{3}]=-i\epsilon_{ab} B_{b}, 
[B_{a},K_{3}]=iB_{a}, [J_{3},K_{3}]=0, 
[S_{1},S_{2}]=0, [S_{a},J_{3}]=-i\epsilon_{ab} S_{b}, 
[S_{a},K_{3}]=-iS_{a}$ where $a,b=1,2$ and $\epsilon_{12}=-\epsilon_{21}=1$. 
Also $[B_{1},P_{1}]=[B_{2},P_{2}]=i P^{+}, [B_{1},P_{2}]=[B_{2},P_{1}]=
0,  [B_{a},P^{-}]=iP_{a},  [B_{a},P^{+}]=0, 
[S_{1},P_{1}]=[S_{2},P_{2}]=i P^{-}, [S_{1},P_{2}]=[S_{2},P_{1}]=
0,  [S_{a},P^{+}]=iP_{a},  [S_{a},P^{-}]=0,  
[B_{1},S_{2}]=- [B_{2},S_{2}]=-iJ_{3}, 
[B_{1},S_{1}]=[B_{2},S_{2}]=-iK_{3} $. 
For $ P_{\mu}=i\partial_{\mu}$,  and 
$M_{\mu\nu}\to L_{\mu\nu}= i(x_{\mu}\partial_{\nu}-x_{\nu}\partial_{\mu})$ 
we find 
$B_{a}=(x^{+}P^{a}-x^{a}P^{+}), S_{a}=(x^{-}P^{a}-x^{a}P^{-}), 
 K_{3}=(x^{-}P^{+}-x^{+}P^{-})$  and 
$  \quad J_{3}=(x^{1}P^{2}-x^{2}P^{1})$. 
Under the conventional {\it parity} operation  ${\cal P}$: 
($\;x^{\pm}\leftrightarrow x^{\mp}, x^{1,2}\to 
-x^{1,2}$) and $(  p^{\pm}\leftrightarrow p^{\mp}, p^{1,2}\to 
-p^{1,2}),$ we find  $\vec J\to \vec J,\, \vec K \to -\vec K $, $B_{a}\to -S_{a}$  
etc..
The  six generators 
$\,P_{l}, \, M_{kl}\,$  leave  $x^{0}=0$ hyperplane 
invariant and are  called [1]   
{\it kinematical}  while the remaining $P_{0},\,M_{0k}$ 
the  {\it dynamical } ones. On the LF 
there are {\it seven}  kinematical generators :  $P^{+},P^{1},
P^{2}, B_{1}, B_{2}, J_{3}$ and  $K_{3} $ which leave the LF hyperplane, 
$x^{0}+x^{3}=0$,  invariant and the three {\it dynamical} 
ones $S_{1},S_{2}$ and  $P^{-}$ 
form a mutually commuting set. We note that each of the set 
 $\{B_{1},B_{2},J_{3}\}$ and   $\{S_{1},S_{2},J_{3}\}$ generates 
an  $E_{2}\simeq SO(2)\otimes T_{2} $ algebra; this will be shown below to be 
relevant for 
defining the {\it spin} for massless particle. Including $K_{3}$ in each set 
we find two subalgebras each with four elements. Some useful identities are 
$e^{i\omega K_{3}}\,P^{\pm}\,e^{-i\omega K_{3}}= e^{\pm \omega}\,P^{\pm},
\, e^{i\omega K_{3}}\,P^{\perp}\,e^{-i\omega K_{3}}= P^{\perp}, 
 e^{i\bar v.\bar B}\,P^{-}\,e^{-i\bar v.\bar B}= 
 P^{-}+\bar v.\bar P + {1\over 2}{\bar v}^{2}P^{+},  
 e^{i\bar v.\bar B}\,P^{+}\,e^{-i\bar v.\bar B}= P^{+}, 
 e^{i\bar v.\bar B}\,P^{\perp}\,e^{-i\bar v.\bar B}= 
 P^{\perp}+v^{\perp} P^{+}, 
 e^{i\bar u.\bar S}\,P^{+}\,e^{-i\bar u.\bar S}= 
 P^{+}+\bar u.\bar P + {1\over 2}{\bar u}^{2}P^{-}, 
  e^{i\bar u.\bar S}\,P^{-}\,e^{-i\bar u.\bar S}= P^{-}, 
 e^{i\bar u.\bar S}\,P^{\perp}\,e^{-i\bar u.\bar S}= 
 P^{\perp}+u^{\perp} P^{-}$ 
  where  $P^{\perp}\equiv\bar P=(P^{1},P^{2}), \, v^{\perp}\equiv {\bar v}
= (v_{1},v_{2})\, $ 
and  $(v^{\perp}. P^{\perp})\equiv(\bar v.\bar P)=v_{1}P^{1}+v_{2}P^{2}$ etc. Analogous expressions with 
$P^{\mu}$  replaced by $X^{\mu}$ can be obtained if we use 
$\,[P^{\mu},X_{\nu}]\equiv [i\partial^{\mu},x_{\nu}]= i\delta^{\mu}_{\nu}\,$.

\bigskip
\nl {\bf Appendix B:\quad  LF Spin Operator. Hadrons in LF Fock
Basis} 
\medskip

The Casimir generators of the Poincar\'e group are : $P^{2}\equiv 
P^{\mu}P_{\mu}$ and  $W^{2}$, where  
$W_{\mu}=(-1/2)\epsilon_{\lambda\rho\nu\mu}
M^{\lambda\rho}P^{\nu}$ defines the  Pauli-Lubanski pseudovector. It follows
from $[W_{\mu},W_{\nu}]=i\epsilon_{\mu\nu\lambda\rho} W^{\lambda}P^{\rho}, 
\quad [W_{\mu},P_{\rho}]=0\;$  and  $\;W.P=0$ that in a   
representation charactarized by particualr eigenvalues of 
the two Casimir operators we 
may simultaneously diagonalize  $P^{\mu}$ along with just one 
component of  $W^{\mu}$. We have 
$ W^{+} =-[J_{3} P^{+}+B_{1} P^{2}-B_{2} P^{1}], 
W^{-} =J_{3} P^{-}+S_{1} P^{2}-S_{2} P^{1}, 
W^{1} =K_{3} P^{2}+ B_{2} P^{-}- S_{2} P^{+},$ and 
$W^{2} =-[K_{3} P^{1}+ B_{1} P^{-}- S_{1} P^{+}]$ and it shows 
that  $W^{+}$ {\it has a  special place} since it 
contains only the kinematical generators [19]. On the LF we define  
  ${\cal J}_{3}= -W^{+}/P^{+}$ as the   {\it spin operator}$^{17}$. 
It may be shown  to 
commute with  $P_{\mu}, B_1,B_2,J_3,$ and  $K_3$. 
For $m\ne 0$ we may use the parametrizations $p^{\mu}:( p^{-}=(m^{2}+
{p^{\perp}}^{2})/(2p^{+}), 
p^{+}=(m/{\sqrt 2})e^{\omega}, 
p^{1}=-v_{1}p^{+},  p^{2}=-v_{2}p^{+})$ and 
${\tilde p}^{\mu}: (1,1,0,0)(m/{\sqrt 2})$ in the rest frame. We have 
$P^{2}(p)= m^{2} I$ and  $W(p)^{2}=
W(\tilde p)^{2}= -m^{2} [J_{1}^2+J_{2}^2+J_{3}^2] = -m^{2} s(s+1) I$ 
where $s$ assumes half-integer values.  
Starting from the rest  state $\vert \tilde p; m,s,\lambda, ..\rangle $
with ${J}_{3}\, \vert\tilde p; m,s,\lambda, ..\rangle
= \lambda \,\vert \tilde p; m,s, \lambda, ..\rangle $ we may build  an 
arbitrary eigenstate of $P^{+}, P^{\perp}, {\cal J}_{3} $  (and $ P^{-}$ ) 
on the LF by 

$$\vert p^{+},p^{\perp}; m,s,\lambda, ..\rangle= e^{i(\bar v. \bar B)} 
e^{-i\omega K_{3}} \vert \tilde p; m,s,\lambda, ..\rangle 
$$ 

\nl If we make use of the following  {\it identity} [19] 
for the  spin operator

$${\cal J}_{3}(p)=\;J_{3}+v_{1}B_{2}-v_{2}B_{1}=\quad 
e^{i(\bar v. \bar B)} \;
J_{3} \;e^{-i(\bar v. \bar B)}  $$

\nl we find ${\cal J}_{3}\, \vert p^{+},p^{\perp}; m,s,\lambda, ..\rangle
= \lambda \,\vert p^{+},p^{\perp};m,s,\lambda, ..\rangle $. 
Introducing  also  ${\cal J}_{a}= 
-({\cal J}_{3}P^{a}+W^{a})/{\sqrt{P^{\mu}P_{\mu}}}$, $a=1,2$, which
do, however, contain 
dynamical generators, we verify that 
$\;[{\cal J}_{i},{\cal J}_{j}]=i\epsilon_{ijk} {\cal J}_{k}$.

For $m=0$ case when $p^{+}\ne0$ 
a convenient parametrization is 
$p^{\mu}:( p^{-}=p^{+} {v^{\perp}}^{2}/2, \,p^{+}, 
p^{1}=-v_{1}p^{+}, p^{2}=-v_{2}p^{+})$ and $\tilde p: 
(0, p^{+}, 0^{\perp})$. We have 
$W^{2}(\tilde p) = -(S_{1}^{2}+S_{2}^{2}){p^{+}}^{2}$ 
and  $[W_{1},W_{2}](\tilde p)=0, \,
[W^{+},W_{1}](\tilde p)=-ip^{+}W_{2}(\tilde p), 
\, [W^{+}, W_{2}](\tilde p)=ip^{+}W_{1}(\tilde p)$ showing that 
$W_{1}, W_{2}$ and  $W^{+}$  generate the algebra 
$SO(2)\otimes T_{2}$. The eigenvalues of 
$W^{2}$ are hence not quantized and they vary continuously. 
This is contrary to the experience so we impose that the physical states 
satisfy in addition  $W_{1,2}
\vert \, \tilde p;\,m=0,..\rangle=0$. Hence 
$ W_{\mu}=-\lambda P_{\mu}$ and  the invariant parameter 
$\lambda $ is taken to define as the {\it spin} of the massless particle.  
From  $-W^{+}(\tilde p)/{\tilde p}^{+}=J_{3}$ we conclude that 
$\lambda$ assumes half-integer values as well.  
We note that   $W^{\mu}W_{\mu}=\lambda^{2} P^{\mu}P_{\mu}=0$  and
that the definition of the {\it LF spin operator  appears  unified 
for massless and massive particles}. A parallel discussion based on 
$p^{-}\ne0$ may also be given. 
%the roles of  $W^{+}$ and  $W^{-}$ are  interchanged. 

As an illustration consider the three particle state on the LF with the 
total eigenvalues 
$p^{+}$,  $\lambda$  and $p^{\perp}$. In the {\it standard frame } 
with  $p^{\perp}=0\;$ it may be written as 
($\vert x_{1}p^{+},k^{\perp}_{1}; \lambda_{1}\rangle
\vert x_{2}p^{+},k^{\perp}_{2}; \lambda_{2}\rangle
\vert x_{3}p^{+},k^{\perp}_{3}; \lambda_{3}\rangle$ ) 
with  $\sum_{i=1}^{3} \,x_{i}=1$, $\sum_{i=1}^{3}\,k^{\perp}_{i}=0$, and 
$\lambda=\sum_{i=1}^{3}\,\lambda_{i}$. Aplying 
$e^{-i{(\bar p.\bar B)/p^{+}}}$ on it we obtain 
($\vert x_{1}p^{+},k^{\perp}_{1}+x_{1}p^{\perp}; \lambda_{1}\rangle
\vert x_{2}p^{+},k^{\perp}_{2}+x_{2}p^{\perp}; \lambda_{2}\rangle
\vert x_{3}p^{+},k^{\perp}_{3}+x_{3}p^{\perp}; \lambda_{3}\rangle $ ) 
now with  $p^{\perp}\ne0$. The  $x_{i}$ and $k^{\perp}_{i}$ indicate 
relative (invariant) parameters and do not depend upon the 
reference frame. The  $x_{i}$ is 
the fraction of the total longitudinal momentum 
carried by the  $i^{th} $ particle while 
 $k^{\perp}_{i}$ its transverse momentum. The state of a pion with 
 momentum ($p^{+},p^{\perp}$),  for example, 
 may be expressed  as 
an expansion over the LF Fock states constituted by the different 
number of partons [7]  

$$\vert \pi : p^{+},p^{\perp} \rangle=
{\sum}_{n,\lambda}\int {\bar \Pi}_{i}{{dx_{i}d^{2}{k^{\perp}}_{i}}\over
{{\sqrt{x_{i}}\,16\pi^{3}}}} \vert n:\,x_{i}p^{+},x_{i}p^{\perp}+
{k^{\perp}}_{i},
\lambda_{i}\rangle\;\psi_{n/\pi}(x_{1},{k^{\perp}}_{1},
\lambda_{1}; x_{2},...) $$

\nl where the summation is over all the Fock states 
$n$ and spin projections  $\lambda_{i}$, with 
${\bar\Pi}_{i}dx_{i}={\Pi}_{i} dx_{i}\; \delta(\sum x_{i}-1), $ 
and ${\bar\Pi}_{i}d^{2}k^{\perp}_{i}={\Pi}_{i} dk^{\perp}_{i} \;
 \delta^{2}(\sum k^{\perp}_{i})$.  The wave function of the 
parton $\psi_{n/\pi}(x,k^{\perp})$ 
indicates the probability amplitue for finding inside the pion 
the partons in the Fock state $n$ carrying 
the 3-momenta 
$(x_{i}p^{+}, x_{i}p^{\perp}+ k^{\perp}_{i}) $. The 
  Fock state of the pion is also 
        off the energy shell : $\,\sum k^{-}_{i} > p^{-}$. 
%Cada component
%de Fock descreve um sistema de partons livres com uma massa cinem\'atica 
%invariante ${\cal M}^{2}={\sum}_{i}^{n} ({k^{\perp}_{i}}^{2}+{m_{i}}^{2})/
%x_{i}$

The {\it discrete symmetry} transformations may also be defined on the 
LF Fock states [19]. 
For example, under the conventional parity ${\cal P} $ 
the  spin operator  ${\cal J}_{3}$ is not left invariant. 
We may rectify this by defining {\it LF Parity operation} by 
${\cal P}^{lf}=e^{-i\pi J_{1}}{\cal P}$. We find 
then  $B_{1}\to -B_{1}, B_{2}\to B_{2}, P^{\pm}\to P^{\pm}, P^{1}\to 
-P^{1}, P^{2}\to P^{2}$ etc. such that 
${\cal P}^{lf}\vert p^{+},p^{\perp}; m,s,\lambda, ..\rangle
\simeq \vert p^{+},-p^{1}, p^{2}; m,s,\,-\lambda, ..\rangle $. Similar
considerations apply for charge conjugation and  time inversion. For example, 
it is straightforward to construct  the free {\it LF Dirac spinor}  
$\chi(p)= 
[\sqrt{2}p^{+}\Lambda^{+}+(m-\gamma^{a}p^{a})\,\Lambda^{-}]\tilde \chi/
{ {\sqrt{\sqrt {2}p^{+}m}}}$ which is also an eigenstate os 
${\cal J}_{3}$ with eigenvalues 
$\pm 1/2$. Here $\Lambda^{\pm}= \gamma^{0}\gamma^{\pm}/{\sqrt 2}=
\gamma^{\mp}\gamma^{\pm}/2=({\Lambda^{\pm}})^{\dagger}$, 
$ (\Lambda^{\pm})^{2}=\Lambda^{\pm}$,  
and $\chi(\tilde p)\equiv \tilde \chi\,$ with  $\gamma^{0}
\tilde \chi= \tilde \chi$. The conventional (equal-time) 
spinor can also be constructed by the  procedure analogous to that 
followed for the LF spinor and it has  the well known form 
$ \chi_{con}(p)=  (m+\gamma.p)\tilde \chi/
{\sqrt{2m(p^{0}+m)}}$. 
Under the conventional parity operation  ${\cal P}: 
\chi'(p')=c \gamma^{0} \chi(p)$ ( since we must require 
$\gamma^{\mu}={L^{\mu}}_{\nu}\,S(L)\gamma^{\nu}{S^{-1}}(L)$ etc. ). We find 
$\chi'(p)=c 
[\sqrt{2}p^{-}\Lambda^{-}+(m-\gamma^{a}p^{a})\,\Lambda^{+}]\,\tilde \chi
/{\sqrt{\sqrt {2}p^{-}m}}$. For $p\neq\tilde p$ 
it is not proportional to $\chi(p)$ in contrast to the result in 
the case of the usual spinor where 
$\gamma^{0}\chi_{con}(p^{0},-\vec p)=\chi_{con}(p)$ for  $E>0$ (and  
$\gamma^{0}\eta_{con}(p^{0},-\vec p)=-\eta_{con}(p)$ for  $E<0$). 
However, applying parity operator twice we do show 
$\chi''(p)=c^{2}\chi(p)$ hence leading to the usual result 
$c^{2}=\pm 1$. The LF parity operator over spin $1/2$ Dirac spinor is 
${\cal P}^{lf}= c \,(2J_{1})\,\gamma^{0}$ and the corresponding transform 
of $\chi$ is shown to be an  eigenstate of ${\cal J}_{3}$. 

\bigskip

\nl {\bf Appendix C: \quad SSB  Mechanism. Continuum Limit of Discretized 
LF Quantized Theory. Nonlocality of LF Hamiltonian}. 
\medskip

The existence of the  continuum limit [20,2] of the 
Discretized Light Cone Quantized 
(DLCQ) [21] theory, the nonlocal nature of the LF Hamiltonian, and the 
description of the SSB on the LF were 
clarified [13] only recently. 

 Consider first the two 
dimensional case  with 
${\cal L}= \; \lbrack
{\dot\phi}{\phi^\prime}-V(\phi)\rbrack$. 
% Here $\tau\equiv x^{+}=(x^0+x^1)/{\sqrt2}$,
%  $x\equiv x^{-}=(x^0-x^1)/{\sqrt2}$, $\partial_{\tau}\phi=\dot\phi , 
 %       \partial_{x}\phi={\phi}'$, and $d^2x=d\tau dx$. 
The eq. of motion, 
$\,\dot{\phi^\prime}=(-1/2)\delta V(\phi)/\delta \phi $, shows that 
$\phi=const. $ is a possible solution. 
We write  
$\;\phi(x,\tau)=\omega(\tau)+\varphi(x,\tau)\;$ 
%where $\omega(\tau)$ 
%corresponds to the {\it bosonic condensate} and $\varphi(\tau,x)$ describes 
%(quantum) {\it fluctuations} above it. 
The general case of 
$\omega=\omega(\tau)$ can be made [2] but to make the discussion here
short we  assume that 
%The value of   $\omega(=\langle 0\vert \phi\vert 0\rangle)$
%will be seen to characterize the corresponding vacuum  state. 
%The translational  invariance of the ground state requires 
$\omega $ is a constant    so that 
${\cal L}={\dot\varphi}{\varphi}^\prime-V(\phi)$.  
Dirac procedure is applied now  to 
construct  Hamiltonian field theory which may be quantized. 
We may avoid using 
distribuitions if we   restrict $x$ to a finite interval from  
 $\,-R/2\,$ to $\,R/2\,$. 
 The 
 {\it physical  limit to the continuum ($R\to\infty\,$)}, however, 
  must be taken latter to remove the spurious finite volume effects. 
Expanding $\varphi$ by Fourier series we obtain 
        $\phi(\tau,x) \equiv \omega +\varphi(\tau,x)
= \omega+{1\over\sqrt{R}} {q_{0}(\tau)}+
{1\over\sqrt{R}}\;{{\sum}'_{n\ne 0}}\;\;
{q}_n(\tau)\;e^{-ik_n x}$ 
 where $\,k_n=n(2\pi/R)$, 
$\,n=0,\pm 1,\pm 2, ...\,$ and 
the {\it discretized theory} Lagrangian becomes 
$\;i{{\sum}_{n}}\;k_n \,{q}_{-n}\;{{\dot q}}_{n}-
                                  \int dx \; V(\phi)$. 
The momenta conjugate to ${q}_n$ are 
 ${p}_n=ik_n{q}_{-n}$ 
 and the canonical LF Hamiltonian is found to be  $\int \,
 dx \,V(\omega+\varphi(\tau,x))$. 
The primary constraints$^{}$ are thus 
${p}_0 \approx 0\;$ and $\;{\Phi}_n \equiv 
\;{p}_n-ik_n{q}_{-n}\approx 0\;$ for $\,n\ne 0\,$. 
We follow the {\it standard}  Dirac procedure [3] and find 
three   weak constraints  $\,p_{0}\approx 0$, $\,\beta
\equiv \int dx \,V'(\phi) \approx 0$, and $\,\Phi_{n}\approx 0\,$ 
for $\,n\ne 0\,$ 
on the phase space and they are shown to be  second class. 
We find for $\,n,m\ne 0:\,$
$\{{\Phi}_n, {p}_0\}\,=\,0,\qquad 
\{{\Phi}_n,{\Phi}_m\}\,=\,-2ik_n \delta_{m+n,0}\;,$
$\quad \{{\Phi}_n,\beta\}\,=\,\{{p}_n,\beta\}\,
=\,-{(1/ \sqrt{R})}\int dx\;\lbrack \,V''(\phi)-V''([{\omega +
q_{0}]/
\sqrt{R}})\,\rbrack \,e^{-ik_nx}\,
\equiv \,-{{\alpha}_{n}/\sqrt{R}},\,$
\quad $\{{p}_0,\beta\}\,=\,-{(1/\sqrt{R})}\int dx\;V''(\phi)\,
\equiv \,-{\alpha/\sqrt{R}},\,$\quad
$\{{p}_0,{p}_0\}\,=\,\{\beta,\beta\,\}\,=\,0\,$. 
Implement first the pair of constraints $\;{p}_0\,\approx 0, 
\,\beta\,\approx 0$ by modifying the Poisson brackets to 
the star  bracket $\{\}^*$  defined by 
$\{f,g\}^*\,=\{f,g\}\,-\lbrack\,\{f,{p}_0\}\;
\{\beta,g\}-\,(p_0\,\leftrightarrow 
\beta)\rbrack\,({\alpha/\sqrt{R}})^{-1}$.
We may then set ${p}_0\,=0$ 
and $\beta \,=0$ as  strong equalities. 
We find  by inspection that the brackets 
$\{\}^*\,$ of the remaining  variables coincide with the standard 
Poisson brackets except for the ones involving $q_{0}$ and 
$\,p_{n}\,$ ($n\ne0$):\quad 
$\,\{q_0,{p}_n\}^*\,=
\{q_0,{\Phi}_n\}^*\,=-({\alpha^{-1}}{\alpha}_n)\,$.  
For example, if 
$V(\phi)=\,({\lambda/4})\,{(\phi^2-{m^2}/\lambda)}^2\;$, 
$\lambda\ge 0, m\ne 0\,$ 
we find  $\;\{q_0,{p}_n\}^*\;
[\{\,3\lambda\,({\omega+q_{0}/\sqrt{R}})^{2}-m^{2}\,\}R\,+
6\lambda(\omega+q_{0}/{\sqrt R})\int\, dx \varphi +
\,3\lambda \,\int\,dx\,\varphi^{2}\,]= 
-\,3\lambda\,[\,2(\omega+q_{0}/{\sqrt R})\,{\sqrt R}
 q_{-n}
 +\int \,dx\,\varphi^{2} 
\,e^{ -ik_{n}x} \,] $.

Implement next the constraints $\,\Phi_{n}\approx 0\,$ 
with $\,n\neq 0$. We have $ C_{nm}\,=\,\{\Phi_{n},\Phi_{m}\}^*\,=
\,-2ik_{n}\delta_{n+m,0}\,$ 
and its inverse is given by  $\,{C^{-1}}_{nm}\,=\,(1/{2ik_{n}})
\delta_{n+m,0}\;$. The  Dirac bracket which takes care of all 
the constraints is then given by

$$\{f,g\}_{D}\,= \,\{f,g\}^{*}\,-\,{{\sum}'_{n}}\,{1\over {2ik_{n}}}
\{f,\,\Phi_{n}\}^*\,\{\Phi_{-n},\,g\}^* $$  

\nl where we may now in addition write $\,p_{n}\,=\,ik_{n}
q_{-n}\,$. It is easily shown that 
$\{q_0,{q}_0\}_{D}\,=0, 
\{q_0,{p}_n\}_{D}\,=\{q_0,\,ik_{n}
q_{-n}\}_{D}\,={1\over 2}\,\{q_0,{p}_n\}^*,
\{q_{n},p_{m}\}_{D}\,={1\over 2}\delta_{nm}$. 

The  limit to the continuum, $R\to\infty$ is taken as usual:  
$\Delta=2\,({\pi/{R}})\to dk\,,\,k_{n}=n\Delta\to k\,,\,\sqrt{R}\,
q_{-n}\to\,lim_{R\to\infty}
 \int_{-R/2}^{R/2}{dx}\,\varphi(x)\,e^{ik_{n}x}\equiv\,\int_{-\infty}^
{\infty}\,dx\, \varphi(x)\,e^{ikx}\,=\,\sqrt{2\pi}{\tilde\varphi(k)}$ 
for all $\,n $,  
$\,\sqrt{2\pi}
\varphi(x)\,=\int_{-\infty}^{\infty}\,dk\,\tilde\varphi(k)\,e^{-ikx}\;$, 
and  $(q_{0}/{\sqrt{R}})\to 0\,$. 
From $\,\{\sqrt{R}q_{m},\sqrt{R}q_{-n}\}_{D}\,=\,
R\,\delta_{nm}/({2ik_{n}})\,$ following from $\{q_{n},p_{m}\}_{D}$ 
for $n,m\ne 0$  
we derive, on using  $\,R\delta_{nm}\to \int_{-\infty}^{\infty}dx e^{i(k-k')x}=
\,{2\pi\delta(k-k')}$, that 
$\{\tilde\varphi(k),\tilde\varphi(-k')\}_{D}\,=\,\delta
(k-k')/(2ik)\,$ 
where  $ k,k'\,\ne 0$. 
If we use the integral representation of the sgn 
function  the well known LF Dirac bracket 
$\{\varphi(x,\tau),\varphi(y,\tau)\}_{D}=-{1\over 4}\epsilon(x-y)$ is
obtained. 
The expressions of  $\{q_{0},p_{n}\}_{D}$ (or $\{q_{0},\varphi'\}_{D}$) 
show that the DLCQ is harder to work with here. 
The  continuum limit of the {\it constraint eq.} $\beta=0$ is 

$$ \eqalign {& \,lim_{R\to\infty} 
{1\over R}\int_{-R/2}^{R/2} dx \,V'(\phi)\equiv  \cr 
& \omega(\lambda\omega^2-m^2)+lim_{R\to\infty} {1\over R}
 \int_{-R/2}^{R/2} dx \Bigl[ \,(3\lambda\omega^2-m^2)\varphi + 
 \lambda (3\omega\varphi^2+\varphi^3 ) \,
\Bigr]=0}$$

\nl while that for the LF Hamiltonian is ($P^{-}\equiv H^{l.f.}$)

$${ P^{-}\,=\int  dx \,\Bigl [\omega(\lambda\omega^2-m^2)\varphi+
{1\over 2}(3\lambda\omega^2-m^2)\varphi^2+
\lambda\omega\varphi^3+{\lambda\over 4}\varphi^4 
\Bigr ]\,}$$

\nl These results  follow  immediately 
if we worked directly  in the continuum formulation; we do have 
to handle generalized functions now.  In the LF Hamiltonian theory 
we have an additional new ingredient in the form of the 
{\it constraint equation}.  Elimination of 
$\omega $ using it would lead to a {\it nonlocal LF Hamiltonian} 
corresponding to the 
local one in the equal-time formulation.  
At the tree or classical level the integrals appearing in in 
the constraint eq.  
are convergent and when  $R\to\infty$  it leads to 
$V'(\omega)=0$. In 
equal-time theory this is essentially {\it added} to  it 
as an  external constraint based on physical considerations. In 
the renormalized theory [13]  the constraint equation 
 describes 
the high order quantum corrections to the tree level value of the condensate.

The quantization is performed via the correspondence$^{}$  
$i\{f,g\}_{D}\to [f,g]$. Hence 
$\varphi(x,\tau)= {(1/{\sqrt{2\pi}})}\int dk\; 
{\theta(k)}\;
[a(k,\tau)$ $e^{-ikx}+{a^{\dag}}(k,\tau)e^{ikx}]/(\sqrt {2k})$, 
were $a(k,\tau)$ and ${a^{\dag}}(k,\tau)$ 
satisfy the canonical equal-$\tau$ commutation relations, 
$[a(k,\tau),{a(k^\prime,\tau)}^{\dag}]=\delta(k-k^\prime)$ etc.. 
The vacuum state is defined 
by  $\,a(k,\tau){\vert vac\rangle}=0\,$, 
$k> 0$ and the tree level  
description of the {\it SSB} is  given as
follows. The values of $\omega=
\,{\langle\vert \phi\vert\rangle}_{vac}\;$ obtained from  
$V'(\omega)=0$ 
the different   vacua in the theory. 
Distinct  Fock spaces corresponding to different values of $\omega$ 
are built as usual by applying the creation operators on the corresponding
vacuum state.  The $\omega=0$ corresponds to a {\it symmetric phase} 
since the Hamiltonian is then symmetric under 
$\varphi\to -\varphi$. For $\omega\ne0$ this symmetry is 
violated and the  system is 
in a {\it broken or asymmetric phase}.  

The extension to $3+1$ dimensions and to global continuous symmetry 
is straightforward [13,2]. Consider  real scalar fields 
$\phi_{a} (a=1,2,..N)\,$ which form  an isovector of global 
internal symmetry group 
 $O(N)$. We now  write 
 $\phi_{a}(x,x^{\perp},\tau)
=\omega_{a}+\varphi_{a}(x,x^{\perp},\tau)$ and 
the Lagrangian density is  ${\cal L}=[{\dot\varphi_{a}}{\varphi'_{a}}-
{(1/ 2)}(\partial_i\varphi_{a})(\partial_i\varphi_{a})-V(\phi)]$,  
where $i=1,2$ indicate the transverse space directions. The Taylor series 
expansion of the constraint equations $\beta_{a}=0$ gives a set of coupled
eqs. $R\,V'_a(\omega)+
\,V''_{ab}(\omega)\int dx \varphi_{b}+\,
V'''_{abc}(\omega)\int dx \varphi_b\varphi_c/2+...=0$. Its discussion at 
the tree level leads to the conventional theory results. 
The  LF symmetry generators are found to be 
$G_{\alpha}(\tau)=
-i\int d^{2} {x^{\perp}} dx \varphi'_{c}(t_{\alpha})_{cd}\varphi_{d} 
=\,\int d^{2}{k^{\perp}}\,dk \, \theta(k)\, 
{a_{c}(k,{k^{\perp}})
^{\dag}} (t_{\alpha})_{cd} a_{d}(k,{k^{\perp}})$ where 
$\alpha,\beta=1,2,..,N(N-1)/2\, $,  are the group indices, 
$t_{\alpha}$ are hermitian and antisymmetric generators of $O(N)$, and 
${a_{c}(k,{k^{\perp}})^{\dag}}$ ($ a_{c}(k,{k^{\perp}})$) is creation (
destruction)  operator  contained in the momentum space expansion 
of $\varphi_{c}$. These  are to be contrasted with the generators 
in the equal-time theory,  
$ Q_{\alpha}(x^{0})=\int d^{3}x \, J^{0}
=-i\int d^{3}x (\partial_{0}\varphi_{a})(t_{\alpha})_{ab}\varphi_{b} 
-i(t_{\alpha}\omega)_{a}\int d^{3}x 
({{d\varphi_{a}}/ dx_{0}})$.  
All  the symmetry generators thus  
annihilate the LF vacuum  and the SSB is  
now seen  in the 
broken symmetry of the quantized theory Hamiltonian. The criterian for 
the counting of the number of Goldstone bosons on the LF is found 
to be the same as in the conventional theory. 
In contrast, the first term 
on the right hand side of $Q_{\alpha}(x^{0})$ does  annihilate  
the conventional theory vacuum  but 
the second term  gives now  non-vanishing contributions  
for some of the (broken) generators. The  symmetry of the
conventional theory vacuum 
is thereby broken  while the quantum Hamiltonian remains invariant.  
The physical content  of SSB in the {\it instant form}
and the {\it front form}, however, is the same 
though achieved  by differnt descriptions. Alternative proofs  
 on the LF,  
in two dimensions, can be given  
of the Coleman's theorem related to the absence of 
Goldstone bosons and  of the pathological nature of 
massless scalar theory; we are unable to implement 
the second class constraints over the phase space.

We remark that the simplicity of the LF vacuum is in a sense 
compensated by the involved nonlocal Hamiltonian. The latter, however, 
may be treatable using advance computational techniques. In a recent 
work [13] it was also shown that renormalized theory may be constructed 
without the need of first solving the constraint eq.  for $\omega$.  
Instead we  may perform  renormalization and 
obtain a renormalized constraint equation.

\bigskip
\nl {\bf Appendix D: \quad Commutators for equal-$x^{-}$}
\medskip
The LF formulation is symmetrical with respect to $x^{+}$ and $x^{-}$ 
and it is a matter of convention that we take the plus component as
the LF {\sl time} while the other as a spatial coordinate. 
The  theory quantized at $x^{+}=const.$  hyperplanes, however, does 
seem to already incorporate in it the information from 
the equal-$x^{-}$ quantized theory. 

For illustration we consider the two dimensional massive free scalar
theory. The LF quantization, assuming 
$x^{+}$ as the LF {\sl time} coordinates, leads  to 
$\omega=0$ and the equal-${x^{+}}$ commutator 
$[\varphi(x^{+},x^{-}),
\varphi(x^{+},y^{-}]=-i\epsilon(x^{-}-y^{-})/4$. The commutator 
can be realized in the momentum space through the expansion (Appendix
C) 

$$\varphi(x^{+},x^{-})={1\over{\sqrt{2\pi}}}\int_{k^{+}>0}^{\infty}\,
 {dk^{+}\over{\sqrt{2k^{+}}}}
\Bigl [ a(k^{+}) e^{-i(k^{+}x^{-}+k^{-}x^{+})} +
        a^{\dag}(k^{+}) e^{i(k^{+}x^{-}+k^{-}x^{+})} \Bigr ]$$

\nl where $[a(k^{+}),{a(l^{+})}^{\dag}]=\delta(k^{+}-l^{+})$ etc. and
$2k^{+}k^{-}=m^{2}$.  It is then easy to show 

$$[\varphi(x^{+},x^{-}),
\varphi(y^{+},x^{-})]={1\over{2\pi}}\int_{k^{+}>0}^{\infty}
{dk^{+}\over {2k^{+}}}\Bigl [e^{ik^{-}(y^{+}-x^{+})}-
e^{-ik^{-}(y^{+}-x^{+})} \Bigr ].$$

\nl We may change the integration variable to $k^{-}$ by 
making use of $k^{-}dk^{+}+k^{+}dk^{-}=0$. Hence on employing the
integral representation 
$\epsilon(x)= (i/\pi){\cal P}\int_{-\infty}^{\infty} (d\lambda/\lambda) \;
exp(-i\lambda x)\,$ we arrive at the equal-$x^{-}$ commutator 

$$[\varphi(x^{+},x^{-}),
\varphi(y^{+},x^{-})]= -{i\over 4} \epsilon(x^{+}-y^{+})$$

\nl The above field expansion on the LF, in 
contrast to the equal-time case, does not 
involve the mass parameter $m$ and 
the same result follows in the massless case also  if we 
assume that $k^{+}=l^{+}$ implies $k^{-}=l^{-}$. 
Defining  the right  and the left movers  by 
$\varphi(0,x^{-})\equiv \varphi^{R}(x^{-})$, 
and $\varphi(x^{+},0)\equiv \varphi^{L}(x^{+})$ we obtain  
$[\varphi^{R}(x^{-}),\varphi^{R}(y^{-})]=(-i/4)\epsilon(x^{-}-y^{-})$ while 
$[\varphi^{L}(x^{+}),\varphi^{L}(y^{+})]=
(-i/4)\epsilon(x^{+}-y^{+})$.  

\eject
\bigskip

\bigskip
\nl {\bf References}
\bigskip
\item{[1]} P.A.M. Dirac, Rev. Mod. Phys. {\bf 21} (1949) 392.

\item{[2]} See for example, {\it Lectures on 
  Light-front quantized field theory:}  
{\sl Spontaneous symmetry breaking. Phase transition in $\phi^{4}$ theory}, 
Proc.  {\it XIV Encontro Nacional de Part\|culas e Campos, Caxambu, MG},  
  pgs. 154-192, {\sl  
Sociedade Brasileira de F\|sica}, Brasil, 1993 and the references
cited therein (
{\rm hep-th@xxx.lanl.gov}/ 9312064). 
%See also 9412204, 9412205. 
\item{[3]}P.A.M. Dirac, {\it Lectures 
in Quantum Mechanics}, Benjamin, New York, 1964; E.C.G. Sudarshan and 
N. Mukunda, {\it Classical Dynamics: a modern perspective}, Wiley, N.Y., 
1974; A. Hanson, T. Regge and C. Teitelboim, {\it Constrained 
Hamiltonian Systems}, Acc. Naz. dei Lincei, Roma, 1976.
\item{[4]} J. Schwinger, Phys. Rev. {\bf 128}, 2425 (1962);
{\it Theoretical Physics}, I.A.E.A, Vienna, p. 89, 1963. 

\item{[5]}S. Weinberg, Phys. Rev. {\bf 150} (1966) 1313. 
\item{[6]}J.B. Kogut and D.E. Soper, Phys. Rev. {\bf D 1} (1970) 2901.

\item{[7]}S.J. Brodsky and H.C. Pauli, {\it Light-cone Quantization and 
QCD}, Lecture Notes in Physics, vol. 396, eds., H. Mitter et. al., 
Springer-Verlag, Berlin, 1991; 
\item{} S.J. Brodsky and G.P. Lepage, 
 in {\it Perturbative Quantum
Chromodynamics}, ed., A.H. Mueller, World Scientific, Singapore, 1989.
\item{[8]} K.G. Wilson, Nucl. Phys. B (proc. Suppl.) {\bf 17} (1990). 
R.J. Perry, A. Harindranath, and K.G. Wilson, 
Phys. Rev. Lett. {\bf 65} (1990)  2959. 
\item{[9]} R.J. Perry, {\it Hamiltonian Light-Front Field Theory and QCD},  
{\sl Hadron Physics 94}, pg. 120, eds., 
 V. Herscovitz et. al., World Scientific, Singapore, 1995.
% and the  references contained therein. 
\item{[10]} J. Lowenstein and J. Swieca, Ann. Phys. (N.Y.) {\bf 68},
172 (1971). 

\item{[11]} See for example, E. Abdalla, M.C. Abdalla and K. Rothe, {\it
Non-Perturbative Methods in Two Dimensional Quantum Field Theory}, 
World Scientific, Singapore, 1991 and the references cited therein.  
\item{[12]} J. Kogut and L. Susskind, Phys. Rev. {\bf D11}, 3594
(1975) and the refs. contained therein.
%A. Casher, J. Kogut, and L. Susskind, Phys. Rev. {\bf D10},
%732 (1974);  J. Kogut and D.K. Sinclair, Phys. Rev. {\bf D10}, 4181 (1974);
\item{[13]}  P.P. Srivastava, {\it On spontaneous symmetry breaking 
mechanism in light-front quantized 
field theory}, Ohio State Univ. preprint 91-0481, {\sl SLAC} database no. 
PPF-9148, November 1991; 
\item{} {\it Light-front Quantization and SSB},  
{\sl Hadron Physics 94}, pg. 253; {\it Phase Transition in Scalar
Theory Quantized on the Light-Front}, ibid, pg.257, eds., 
 V. Herscovitz et. al., World Scientific, Singapore, 1995 (
 hep-th@xxx.lanl.gov. no. 9412204 and 205;  Nuovo Cimento {\bf A107}
(1994) 549; {\bf A108} 35 (1995). 
\item{[14]}G. McCartor, Z. Phys. C {\bf 64}, 349 (1994) and the refs.
cited therein; 
\item{} J.P. Vary, T.J. Fields, and H.J. Pirner, Phys. Rev. {\bf 
D53}, 7231 (1996).  

\item{[15]} D. Matis and Lieb, J. Math. Phys. {\bf 6}, 304 (1965); 
S. Coleman, Phys. Rev. {\bf D11}, 2088 (1975); 
I. Affleck, Nucl. Phys. {\bf B265} [FS15], 409 (1986). See also [12].
\item{[16]} See for example, G. 't Hooft, in {\it New Phenomena in Subnuclear
Physics}, ed. A. Zichichi, vol 13A (Plenum, New York, 1977). 

\item{[17]} E. Abdalla and M.C. Abdalla, Int. Jour. Mod. Phys. 
{\bf A10}, 1611 (1995).
\item{[18]} {\it work in progress}. 

\item{[19]} P.P. Srivastava, {\sl Lightfront quantization of 
field theory} in {\it Topics in Theoretical Physics}, 
{\sl Festschrift for Paulo Leal Ferreira}, eds., V.C. Aguilera-Navarro et.
al., pgs. 206-217,  IFT-S\~ao Paulo, SP, Brasil (1995); 
hep-th@xxx.lanl.gov/9610044. 

\item{[20]} P.P. Srivastava, {\it Spontaneous symmetry breaking mechanism 
in light-front quantized field theory} - {\it Discretized formulation},  
 Ohio State University preprint  92-0173, {\sl SLAC} PPF-9222, April 1992, 
 Contributed 
to {\it XXVI Intl. Conference  on High Energy Physics, Dallas, Texas},  
August 92, (AIP Conference Proceedings 272, pg. 2125 Ed. J.R. Sanford ), 
hep-th@xxx.lanl.gov/ 9412193.

\item{[21]} H.C. Pauli and S.J. Brodsky, Phys. Rev. {\bf D32} (1985) 1993.

\bye